\documentstyle[12pt]{article}
\begin{document}
\begin{flushright}
{BIHEP-TH-95-29}
\end{flushright}

\begin{center}
 {\large\bf	SU(3) Flavor Dependence of the Heavy Quark Symmetry\footnote
{\small\sl  Invited talk given by T.Huang at International Symposium on
Particle Theory and Phenomenology, Ames, Iowa, 22-24 May 1995}}
\end{center}
\vspace{0.5cm}
\begin{center}
{\bf Tao  Huang  and  Chuan-Wang Luo}\\
{\small\sl Institute of High Energy Physics, P.O.Box 918(4), Beijing 100039,
China}
\end{center}
\vspace{0.6cm}

\begin{abstract}
	SU(3) flavor dependence of the leading and
subleading parameters appeared in the heavy quark expansion of the heavy-light
mesons are systematically analyzed by using QCD sum rules.
\end{abstract}

\small
\begin{flushleft}
{\Large\bf I $~$ Introduction} \\
\end{flushleft}

	As the heavy quark goes into the
infinite mass limit, the theory of strong interaction-QCD exhibits a new
spin-flavor symmetry [1],i.e., the so-called heavy quark symmetry
[1,2]. This new
symmetry leads to many remarkable relations among the hadronic matrix elements
between mesons with different spin and flavor, and reduced them to several
independent universal functions[1,3,4].
These universal functions represent the nonperturbative dynamics of the
weak decays of the heavy-light mesons.
 Therefore, to study  them becomes very necessary, which would not only
provide the clear magnitude of them and
enlarge the predictive power of the heavy quark effective theory(HQET),
but also would give us a better
understanding of the nonperturbative nature of the strong interaction,
especially would provide an improvement of our conventional nonperturbative
approaches.
In this talk, we present  QCD sum rule analysis to both strange and non-strange
heavy-light mesons to discuss the important property
of the heavy quark symmetry---its SU(3) flavor dependence.
Our investigation includes decay constants, masses as well as weak form
factors.

 \begin{flushleft}
 {\large\bf II.$~$ Heavy quark symmetry at the leading order}
 \end {flushleft}

       In HQET, the low energy parameter $F_a(\mu)$ of heavy meson
$M_a(\bar{q}Q)$ is defined by [4]
\begin{equation}
 <0|\bar{q}\Gamma h_{Q}|M_a(v)>\;=\;\frac{F_a(\mu)}{2} Tr[\Gamma M(v)],
 \end{equation}
where  $M(v)$ is the spin wavefunction of heavy meson $M_a(v)$ in HQET
\begin{equation}
M(v)\;=\;\sqrt{m_{Q}}\frac{1+\not v}{2} (-i\gamma_{5}).
\end{equation}
For convenient, from now on,  we will omit all light flavor index a=u,d,s
except in the numerical analysis.

Starting from the two-point correlation function in HQET,
\begin{equation}
    \pi_5(\omega)\;=\;i\int d^4 x e^{i k\cdot x}<0|T{A^{(v)}_5(x),
A^{(v)+}_5(0)}|0>,
\end{equation}
where $A_5^{(v)}=\bar{q}\gamma_{5} h_{Q}$ and $\omega=2 k\cdot v$, one obtains
the sum rule for $F(\mu)$[5,6]

\begin{eqnarray}
  F^2(\mu) e^{-2\bar{\Lambda}/T} & = &\frac{3}{8\pi^2}\int^{\omega^c}_{
2m_q}ds \sqrt{s^2-4m^2_q}\{[2 m_q + s]\cdot[1 + \frac{\alpha_s(\mu)}{\pi}
(2\ln\frac{\mu}{s}+\frac{4}{9}\pi^2+\frac{17}{3})] \nonumber \\
&+& m_q \frac{\alpha_s(\mu)}{\pi}[2\ln\frac{\mu}{s}+\frac{7}{3}]\}e^{-s/ T}
 \nonumber  \\
& &  -<0|\bar{q}q|0>[1-\frac{m_q}{2T}+\frac{m^2_q}{2T^2}+
\frac{2\alpha_s(\mu)}{3\pi}]
-\frac{<0|\frac{\alpha_s}{\pi}GG|0>m_q}{4T^2}[\gamma-0.5-ln\frac{T}{\mu}]
   \nonumber   \\
& & +\frac{g_s <0|\bar{q}\sigma Gq|0>}{4T^2}+ \frac{4\pi\alpha_s}{81T^3}
<0|\bar{q}q|0>^2  \\
&=& J_F(\omega^c, T),
\end{eqnarray}
with $\gamma=0.5772 $ being the Euler constant. Taking the derivative with
respect to the inverse of T ,  one can  obtain the sum rule for
$\bar{\Lambda}_a$.

	The Isgur-Wise function $\xi(v\cdot v^{\prime},\mu)$ is defined
 by the matrix element at the leading order in $\frac{1}{m_Q}$[7]
\begin{equation}
<M(v^{\prime})|\bar{h}_{Q_2}(v^{\prime})\Gamma h_{Q_1}(v)|M(v)>\;=\;
-\xi(v\cdot v^{\prime}, \mu) Tr[\bar{M}(v^{\prime})\Gamma M(v)].
\end{equation}
Similar to the above approach, it is not difficult to get the sum rule for
the Isgur-Wise function
\begin{equation}
\xi(y,\mu)\;=\;\frac{K(T,\omega^c,y)}{K(T,\omega^c,1)}.
\end{equation}
At the lowest order, $K(T,\omega^c,y)$ is given by [5]
\begin{eqnarray}
K(T,\omega^c,y) = & \frac{3}{8\pi^2}(\frac{2}{1+y})^2  \int^{\omega^c}_{m_q
\sqrt{2(1+y)}}   d\alpha [\alpha+(1+y)m_q]\sqrt{\alpha^2-2(1+y)m^2_q}
 e^{-\alpha/T}   \nonumber   \\
 &  -<0|\bar{q}q|0>[1-\frac{m_q}{2T}+\frac{m^2_q}{4T^2}(1+y)]  \nonumber  \\
 & + <0|\frac{\alpha_s}{\pi}GG|0>[\frac{y-1}{48T(1+y)}-\frac{m_q}{4T^2}
(\gamma-0.5-\ln\frac{T}{\mu})]  \nonumber   \\
 & + \frac{g_s<0|\bar{q}\sigma Gq|0>}{4T^2}\frac{2y+1}{3}+
\frac{4\pi\alpha_s <0|\bar{q}q|0>^2}{81T^3} y.
\end{eqnarray}
However,  radiative correction in $K(T,\omega^c,y)$ is very complex
and will be presented  in Ref.[8].

	In the numerical analysis of sum rules,	we take the parameters
such as condensates and $m_q$ as in [4-7] and set the scale $\mu=1GeV$.

	Evaluations of sum rules for $F_a$ and $\bar{\Lambda}_a$ give
\begin{eqnarray}
\bar{\Lambda}_s\simeq 0.66\pm 0.08 GeV,&\hat{F}_s\simeq 0.48\pm 0.08 GeV^{3/2},
\\
\bar{\Lambda}_{u,d}\simeq 0.58\pm 0.08 GeV,&\hat{F}_{u,d}\simeq 0.39
\pm 0.07 GeV^{3/2}.
\end{eqnarray}
where $\hat{F}$ is a renormalization group invariant defined in Ref.[6].

Similarly, one also gets
$\Delta\bar{\Lambda}=\bar{\Lambda}_s-\bar{\Lambda}_{u,d}$
and the ratio $R_F=F_s(\mu)/F_{u,d}(\mu)$ with the corresponding sum rules[6],
\begin{eqnarray}
\Delta\bar{\Lambda}=82\pm 8 MeV, & R_F=1.23\pm 0.03.
\end{eqnarray}

	The numerical analysis [5] show that the Isgur-Wise function
$\xi_a(y)$  depends on the parameters $\omega_a^c$ and T  very weakly.
At the center of the sum rule window  T=0.8GeV,  we find the slope
parameter $\varrho^2_a$ defined as $\varrho^2_a=-\xi^{\prime}_a(y=1,\mu)$
has an important property $\varrho^2_s > \varrho^2_{u,d}$ and
$R_{IW}=\xi_s/\xi_{u,d}\simeq (98.3\pm 0.7 )\% $ at $y=1.6$(for $\sigma
(y)=1$).
Although different continuum model $\sigma(y)$ gives different value for
$R_{IW}$, the property $R_{IW} < 1 $(for $y\not = 1$) is independent
of the model choice.

	In summary ,  It is very interesting to find that
the Isgur-Wise function for $ B_s\rightarrow D_s$ falls faster than the
Isgur-Wise function for $B_{u,d}\rightarrow D_{u,d}$, which is just contrary
to the prediction of the heavy meson chiral perturbation theory where only
SU(3) breaking chiral loops are calculated [9]. Our result $R_{IW}\leq 1$
agrees with that of other calculations [10]. It is expected that the future
experiments can test this result and reveal the underlying mechanism of
 SU(3) breaking effects.

\begin{flushleft}
{\large \bf III. Subleading corrections to the decay constants and the masses}
\end{flushleft}

	In the HQET, the corresponding effective lagrangian is
\begin{equation}
L=\bar{h}_v iv\cdot D h_v + \frac{L_K}{2 m_Q} + \frac{C_{mag}(\mu)
L_S}{2 m_Q}.
\end{equation}
The two operators $L_K$ and $L_S$ have clear physical meaning:
$L_K$ is just the kinetic energy operator of the heavy quark
\begin{equation}
L_K\;=\; \bar{h}_v (iD)^2 h_v,
\end{equation}
while $L_S$ is the corresponding chromomagnetic interaction operator
\begin{equation}
L_S\;=\;\frac{1}{2}\bar{h}_v g_s\sigma_{\mu\nu} G^{\mu\nu} h_v,
\end{equation}
whose  Wilson coefficient $C_{mag}(\mu)$ is given in a hybrid way [2]
\begin{equation}
C_{mag}(\mu)\;=\;Z^{-3/\beta_0}[1 + \frac{13}{6}\frac{\alpha_s}{\pi}]
\end{equation}
with $Z=\alpha_s(\mu)/\alpha_s(m_Q)$ and $\beta_0=11-\frac{2 n_f}{3}$.
Then the decay constants can be expanded up to the $\frac{1}{m_Q}$ order as
\begin{eqnarray}
<0|\bar{q}\Gamma Q|M(v)>&=& [ C_1(\mu)+ \frac{1+d_M}{4}C_2(\mu)] F(\mu)
Tr[\Gamma M(v)] \{ 1+ \frac{1}{m_Q} [ G_K(\mu)+ a(\mu) \frac{m_q}{6} -
\frac{\bar{\Lambda}}{6}b(\mu)] \nonumber    \\
&+& \frac{2 d_M}{m_Q} [ C_{mag}(\mu) G_{\Sigma}
(\mu)+A(\mu) \frac{m_q}{12}-\frac{\bar{\Lambda}}{12}B(\mu) ]\},
\end{eqnarray}
The $\frac{1}{m_Q}$ corrections involve the three universal parameters: $
\bar{\Lambda}$, $G_K$ and $G_{\Sigma}$. Among them, $\bar{\Lambda}$ is the
mass parameter, which measures the mass difference between the meson and heavy
quark in the heavy quark limit. The universal parameters $G_K$ and $G_{\Sigma}$
come from insertions of the subleading operators in the effective lagrangian
into the matrix element of the leading order current $\bar{q}\Gamma
h_v$[4],i.e.,
\begin{eqnarray}
<0|i\int dx L_K(x), \bar{q}(0)\Gamma h_v(0)|M(v)>&=& F(\mu) G_K(\mu) Tr[\Gamma
M(v)],    \\
<0|i\int dx L_{\Sigma}(x), \bar{q} \Gamma h_v(0)|M(v)>&=& 2 d_M F(\mu)
G_{\Sigma}(\mu) Tr[\Gamma M(v)]
\end{eqnarray}
with $d_P=3$ for the pseudoscalar meson, $ d_V=-1$ for the vector meson.

The short distance coefficients  $C_1(\mu)$, $C_2(\mu)$, $b(\mu)$ and $B(\mu)$
  are given in [4,11]. For $m_q \not=0$,
$a(\mu)$ and $A(\mu)$ are nonzero. As $C_{mag}(\mu)$, we  express them
in the hybrid approach in which the scale in the next
-leading corrections is ambiguous for some unknown two-loop anomalous
dimensions.
\begin{eqnarray}
a(\mu)&=&-2+\frac{\alpha_s(m_Q)-\alpha_s(\mu)}{4\pi}\frac{S_{hl}}{2}
-\frac{5\alpha_s(m_Q)}{6\pi}+ x^{-4/\beta_0}[2-\frac{\alpha_s(m_Q)-
\alpha_s(\mu)}{2\pi}S_{hl}+\frac{7\alpha_s(m_Q)}{6\pi}]  \nonumber \\
&-&x^{-3/\beta_0}[\frac{13\alpha_s}{6\pi}-\frac{2\alpha_s(m_Q)}{\pi}
+\frac{\alpha_s(m_Q)-\alpha_s(\mu)}{4\pi}S_{hl}]
+\frac{5\alpha_s}{6\pi}x^{-2/\beta_0},   \\
A(\mu)&=& \frac{3}{8}S_{hl}\frac{\alpha_s(m_Q)-\alpha_s(\mu)}{\pi}
-\frac{13\alpha_s(m_Q)}{6\pi}+\frac{7}{3}x^{-4/\beta_0}[1
-\frac{\alpha_s(m_Q)-\alpha_s(\mu)}{4\pi}S_{hl}+\frac{29\alpha_s(m_Q)}{42\pi}]
\nonumber   \\
&-&\frac{1}{3}x^{-3/\beta_0}[4-\frac{\alpha_s(m_Q)-\alpha_s(\mu)}{\pi}\frac{5
S_{hl}}{4}+\frac{14\alpha_s(m_Q)}{3\pi}-\frac{13\alpha_s}{6\pi}]
-\frac{5\alpha_s}{18\pi}x^{-2/\beta_0},
\end{eqnarray}
with $S_{hl}=3\frac{153-19n_f}{(33-2n_f)^2}-\frac{381+28\pi^2-30n_f}
{36(33-2n_f)}$.

	In order to derive the sum rules for $G_K$ and $G_{\Sigma}$,
the correlation functions are chosen as [6,12]
\begin{eqnarray}
\bar{\pi}_K(\omega)&=&i^2\int dxdy e^{ik\cdot (x-y)}<0|T{[\bar{q}
\Gamma_M h_Q]_x, L_K(0), [\bar{h}_Q\bar{\Gamma}_M q]_y}|0>,  \\
\bar{\pi}_{\Sigma}(\omega)&=& i^2\int dxdy e^{ik\cdot (x-y)}<0|T{[\bar{q}
\Gamma_M h_Q]_x, L_{\Sigma}(0), [\bar{h}_Q\bar{\Gamma}_M q]_y}|0>.
\end{eqnarray}
By using the standard procedure, one can easily obtain the final sum rules
\begin{equation}
G_K(\mu) = \frac{1}{2}\frac{d}{dT}[\frac{T J_K(\omega^c,T)}{J_F(\omega^c,T)}],
\end{equation}
\begin{equation}
G_{\Sigma}(\mu) = \frac{1}{4}\frac{d}{dT}[\frac{T J_{\Sigma}(\omega^c,T)}
{J_F(\omega^c,T)}],
\end{equation}
where $J_K(\omega^c,T)$ and $J_{\Sigma}(\omega^c,T)$ are evaluated up to
two loops [6].  We find that the SU(3) breaking effects in the
decay constant of the pseudoscalar are about $17\%$ for the beauty meson
and $13\%$ for the charmed meson respectively, i.e., $f_{B_s}/f_B=1.17\pm 0.03$
, $f_{D_s}/f_D=1.13\pm 0.03$, in which the SU(3) breaking effects in the
subleading order is about $-3\%$ of the corresponding leading one for the
beauty meson, about $-5.5\%$ for the charmed meson. In addition, the ratio of
 the vector to pseudoscalar meson decay constants are found to be
$f_{B^*_s}/f_{B_s}=f_{B^*}/f_B=1.05\pm 0.02; f_{D^*_s}/f_{D_s}=1.23\pm 0.06,
f_{D^*}/f_D=1.24\pm 0.06$.

	Subleading corrections to the heavy meson masses include two parameters:
the heavy quark kinetic energy parameter K and the chromomagnetic interaction
parameter $\Sigma$, which are  defined respectively by [13]
\begin{equation}
K\;=\;=<M(v)|L_K|M(v)>[ < M(v)|\bar{h}_v h_v|M(v)>]^{-1}.
\end{equation}
and
\begin{equation}
<M(v)|L_{\Sigma}|M(v)>\;=\;-d_M\Sigma(\mu)  < M(v)|\bar{h}_v h_v|M(v)>.
\end{equation}
Luckily, the chromomagnetic interaction parameter $\Sigma$ can be extracted
from the experiment data and has been found to be almost independent of the
light flavor [14]. However, it is significant to find from the QCD sum rules
that the maximum SU(3) breaking  effect in
the heavy quark kinetic energy K is about $4\%$[15], i.e.,
\begin{eqnarray}
K_{u,d}=-0.47\pm 0.10 GeV^2,   & K_s=-0.49\pm 0.11 GeV^2
\end{eqnarray}
and
\begin{equation}
R_K=K_s/K_{u,d}=1.02\pm 0.02.
\end{equation}

\begin{flushleft}
{\large \bf IV. Subleading Isgur-Wise functions}
\end{flushleft}

	In order to get the subleading corrections to the weak form factors,
we should construct the sum rules for the subleading Isgur-Wise functions
$\xi_3$,$\chi_1$,$\chi_2$ and $\chi_3$. Although the strange quark is more
heavier than the up and down quarks, QCD sum rules analysis [16] show that the
$1/m_Q$ corrections to the weak
form factors are small for all light flavors, especially the subleading
Isgur-Wise functions $\chi_2$,$\chi_3$ arisen by the chromomagnetic
interactions are very small. The SU(3) breaking effects in all subleading
Isgur-Wise functions are about $18\sim 20\%$ and almost independent of y.

\begin{flushleft}
{\large \bf V. Summary}
\end{flushleft}

	In this talk, we systematically
discuss SU(3) flavor dependence of the heavy quark symmetry. To be specific,
we evaluate decay constants of the  heavy-light mesons, the fundamental
 mass
observable $\bar{\Lambda}$ in HQET, the heavy quark kinetic energy, Isgur-Wise
function and subleading Isgur-Wise functions and their SU(3) breaking effects.
We find i) the $1/m_Q$ corrections to  decay constants of the heavy-light
mesons, especially for the charmed mesons are large, and for the SU(3) breaking
quantities such as $f_{B_s}/f_B$,$f_{D_s}/f_D$ etc., the $1/m_Q$ corrections
are about $3\%\sim7\%$. ii) for SU(3) breaking effects in the mass, the main
contribution is $\Delta\bar{\Lambda}$, and the heavy quark kinetic energy
almost
is flavor independent. iii) the Isgur-Wise function
in HQET sum rules shows that SU(3) flavor symmetry is a good  approximation but
 there is a discrepancy between result of HQET sum
rules  and that of the heavy meson chiral perturbation theory
on the SU(3) breaking behavior.
SU(3) breaking effects in all subleading Isgur-Wise functions are large and
almost independent of y. iv) the $1/m_Q$ corrections to the weak form factors
are small, especially the subleading form factors $\chi_2$, $\chi_3$ arisen by
the chromomagnetic interactions are very small.

\end{document}